\documentclass[sn-mathphys,Numbered]{sn-jnl}
\usepackage{graphicx}
\usepackage{float}
\usepackage{multirow}%
\usepackage{amsmath,amssymb,amsfonts}
\usepackage{amsthm}
\usepackage{mathrsfs}
\usepackage[title]{appendix}
\usepackage{xcolor}
\usepackage{textcomp}
\usepackage{manyfoot}%
\usepackage{booktabs}%
\usepackage{algorithm}
\usepackage{mathtools}
\usepackage{algorithmicx}%
\usepackage{algpseudocode}%
\usepackage{listings}%
\usepackage{lmodern}
\usepackage[capitalize]{cleveref}
\usepackage{anyfontsize}
\usepackage{hyperref}
\usepackage{cleveref}
\usepackage{mathtools}
\usepackage{subcaption}

\begin{document}

\title{Revisiting Black Hole Embedded in King Dark Matter Halo}

\author[1]{David Senjaya} 
\email{davidsenjaya@protonmail.com} 
\affil[1]{Department of Physics, Faculty of Science, Mahidol University, Bangkok 10400, Thailand}

\maketitle{}

\begin{abstract} 

Motivated by recent criticisms of black hole solutions incorporating dark matter through velocity matching, we construct a black hole spacetime by directly incorporating the dark matter halo density into the Einstein field equations as an energy-density source. We obtain an exact static, spherically symmetric solution and investigate the effects of the dark matter halo on the spacetime geometry, null geodesics, photon sphere, black hole shadow, quasinormal modes, and thermodynamic properties. We show that the halo modifies the trajectories of photons and shifts the radius of the unstable circular photon orbit, thereby altering the photon sphere and the apparent shadow. By relating the Lyapunov exponent of the unstable circular null orbit to the quasinormal frequencies in the eikonal limit, we determine the effects of the halo on the oscillation frequencies and damping rates. We further investigate the thermodynamic properties through the black hole mass, Hawking temperature, entropy, heat capacity, and Gibbs free energy. The results reveal that the dark matter halo modifies the thermodynamic stability and phase structure relative to the Schwarzschild case, giving rise to thermodynamic behavior that is absent in vacuum. Our findings provide a framework for examining the effects of dark matter halos on black hole observables and thermodynamic properties.

\end{abstract}

\section{Introduction}\label{sec: intro}
In recent years, increasing attention has been devoted to black hole models embedded in realistic dark matter environments. In particular, black holes surrounded by dark matter halos described by double power-law density profiles of Dehnen type have been extensively investigated, including the configurations $(1,4,0)$ \cite{Gohain:2024eer}, $\left(1,4,\frac{1}{2}\right)$ \cite{Senjaya:2025via}, $(1,4,1)$ \cite{Toshmatov:2025rln}, $\left(1,4,\frac{3}{2}\right)$ \cite{Senjaya:2026mkl}, $(1,4,2)$ \cite{Uktamov:2025lwb}, and $\left(1,4,\frac{5}{2}\right)$ \cite{Al-Badawi:2024asn}. These studies have demonstrated that dark matter can substantially influence the properties of black holes, including their thermodynamic behavior, gravitational lensing, and shadow structure. Such systems therefore provide a useful framework for investigating the interplay between black holes and their surrounding matter distributions.

Observational advances, including the detection of gravitational waves by the LIGO/Virgo collaborations and the imaging of black hole shadows by the Event Horizon Telescope \cite{LIGOScientific:2016vlm,EventHorizonTelescope:2019ggy}, have provided important tests of General Relativity in the strong-field regime. On galactic scales, however, the observed dynamics cannot be accounted for by baryonic matter alone. In particular, the approximately flat rotation curves of spiral galaxies provide strong evidence for the presence of a non-luminous matter component, commonly referred to as dark matter (DM). Dynamical observations indicate that dark matter constitutes a dominant fraction of the mass inferred from the motion of stars and other tracers in galaxies. Moreover, supermassive black holes are commonly found at the centers of massive galaxies and are expected to reside within extended dark matter halos \cite{Rani:2025esb}. The study of black holes embedded in dark matter environments is therefore relevant to understanding the connection between strong-field gravity and the large-scale structure of galaxies.

The double power-law family of density distributions provides a flexible analytic framework for modeling dark matter halos and stellar systems \cite{Zhao1996,Dehnen1993}. In its general form, the density profile is given by
\begin{equation}
\rho(r) = \rho_0 \left(\frac{r}{r_0}\right)^{-\gamma} \left[1 + \left(\frac{r}{r_0}\right)^\alpha\right]^{\frac{\gamma - \beta}{\alpha}},
\end{equation}
where $(\alpha,\beta,\gamma)$ characterize the sharpness of the transition between the inner and outer regions, the outer logarithmic slope, and the inner logarithmic slope, respectively. By choosing different values of these parameters, a wide variety of commonly used density profiles can be recovered. These include cuspy profiles such as the Navarro--Frenk--White (NFW) profile, $(\alpha,\beta,\gamma)=(1,3,1)$, and the Hernquist profile, $(1,4,1)$, as well as cored profiles such as the pseudo-isothermal profile, $(2,2,0)$, the cored Plummer profile, $\left(2,4,0\right)$ \cite{Senjaya:2026asu}, and the Burkert profile, $\left(\frac{3}{2},3,0\right)$ \cite{Benkrane:2025hpt}. This flexibility makes the double power-law parametrization particularly suitable for studying how different halo structures affect the geometry and physical properties of black holes.

Very recently, a spherically symmetric black hole spacetime surrounded by a King dark matter halo was proposed in Ref.~\cite{Al-Badawi:2026cjx}. However, the spacetime construction presented therein was not obtained as an exact solution of the Einstein field equations for the prescribed dark matter density. Consequently, the resulting metric does not consistently reproduce the assumed dark matter distribution when the corresponding Einstein tensor is evaluated, with the inconsistency being particularly evident in the $G_{00}$ component. This issue raises questions concerning the consistency of the resulting spacetime and, consequently, the physical interpretation of the properties derived from it. The main purpose of the present work is therefore to revisit this problem from first principles and construct the black hole--dark matter spacetime by directly solving the Einstein field equations with the King density profile as the prescribed matter source. This approach ensures consistency between the metric and the corresponding energy--momentum tensor and provides a self-consistent gravitational realization of the King-type dark matter configuration.

We focus on the King profile~\cite{King1962}, which is characterized by a finite-density central core and a smoothly decreasing density at large radii. King-type models were originally introduced in the context of self-gravitating stellar systems and have subsequently been employed phenomenologically in modeling cored dark matter distributions. In contrast to cuspy profiles such as the Navarro--Frenk--White (NFW) and Moore profiles, whose densities increase toward the center, the King profile considered here has a finite central density, corresponding to $\gamma=0$. This cored behavior makes it particularly useful for investigating black holes embedded in dark matter environments in which the central density does not diverge. Such cored halo structures have also been considered in connection with dwarf spheroidal and low-surface-brightness galaxies, where observations have motivated alternatives to cuspy dark matter profiles \cite{deBlok2001}.

For the King model considered in this work, the parameters take the values
\begin{equation}
(\alpha, \beta, \gamma) = (2, 3, 0),
\end{equation}
yielding
\begin{equation}
\rho_{DM}(r) = \rho_0 \left[1 + \left(\frac{r}{r_0}\right)^2\right]^{-3/2} = \frac{\rho_0}{\left[1 + \left(\frac{r}{r_0}\right)^2\right]^{3/2}},
\label{rho}
\end{equation}
which follows directly from the general double power-law parametrization. Here, $\rho_0$ denotes the central density, while $r_0$ is the characteristic length scale associated with the transition from the central core to the outer halo.

The profile in Eq.~\eqref{rho} approaches a constant density at the center and exhibits a power-law decline at large radii. Specifically, for $r\ll r_0$, one has $\rho_{DM}(r)\approx\rho_0$, whereas for $r\gg r_0$,
\begin{equation}
\rho_{DM}(r)\sim \rho_0\left(\frac{r}{r_0}\right)^{-3}.
\end{equation}
The $r^{-3}$ asymptotic behavior leads to a logarithmic growth of the enclosed mass, $M(r)\sim\ln r$, and therefore the total mass does not converge in the formal limit $r\rightarrow\infty$. Thus, although the profile provides a finite central density and a physically useful description over a finite radial range, its analytic extension to arbitrarily large radii requires an outer cutoff. In an astrophysical setting, such a cutoff may be associated with a characteristic halo scale, such as the virial radius. The logarithmic growth of the enclosed mass at large radii should therefore be understood as a consequence of the idealized infinite extension of the analytic profile rather than as a physical prediction for an infinitely extended dark matter halo.

Understanding the influence of dark matter halos on black holes is important for exploring the interplay between compact objects and their galactic environments. In particular, supermassive black holes at galactic centers are embedded in environments containing matter distributed over a wide range of spatial scales. The surrounding dark matter distribution can consequently modify the gravitational field experienced by particles and photons and, in turn, affect observable properties associated with the central black hole. Black hole--dark matter systems therefore provide a natural framework for investigating the connection between strong-field gravitational phenomena, dark matter structure, and galactic dynamics.

The thermodynamic properties of black holes provide an additional perspective on these systems. Following the pioneering works of Bekenstein and Hawking, black holes possess an entropy proportional to the area of their event horizon, while the classical area theorem states that the horizon area cannot decrease under appropriate physical conditions \cite{Hawking:1976de}. In the standard four-dimensional setting, stationary black holes are characterized by a small number of macroscopic parameters, most notably their mass $M$, charge $Q$, and angular momentum $J$ \cite{Israel:1967wq,Carter:1971zc,Bekenstein:1972tm,Bekenstein:1973ur,Izmailov:2019cqr}. Together with the Hawking temperature and entropy \cite{Hawking:1976de}, these quantities form the foundation of black hole thermodynamics and are summarized by the four laws of black hole mechanics. More recently, black hole chemistry has extended this framework by interpreting the cosmological constant as a thermodynamic pressure, leading to rich phase structures and thermodynamic analogies with conventional systems \cite{Mann:2024sru,Mann:2025xrb}. The presence of a surrounding dark matter halo introduces an additional environmental contribution to the spacetime geometry and can consequently modify the thermodynamic properties of the black hole.

This paper is organized as follows. First, we construct a static, spherically symmetric black hole embedded in a King dark matter halo by solving the Einstein field equations with the prescribed density profile. We then analyze the null geodesics of the resulting spacetime and evaluate the Lyapunov exponent associated with the stability of unstable circular null orbits. The quasinormal mode spectrum is subsequently examined to investigate the dynamical response of the black hole to perturbations. Finally, we study the thermodynamic properties of the black hole--dark matter system. These analyses provide a systematic assessment of the effects of the King dark matter profile on the spacetime geometry, photon dynamics, orbital stability, quasinormal modes, and black hole thermodynamics.

\section{Black Hole Construction}

To investigate the gravitational influence of the King dark matter halo on a central compact object, we construct a static, spherically symmetric black hole solution embedded in the surrounding dark matter distribution. We consider the metric ansatz
\begin{gather}
ds^2 = -h(r)dt^2 + \frac{dr^2}{h(r)}
+ r^2\bigl(d\theta^2+\sin^2\theta\,d\phi^2\bigr).
\label{1stmetric}
\end{gather}

The energy--momentum tensor is modeled as an anisotropic fluid,
\begin{equation}
T^\mu_{\ \nu}
=
\mathrm{diag}\left[-\rho_{\rm DM}(r),p_r(r),p_t(r),p_t(r)\right],
\end{equation}
where $\rho_{\rm DM}(r)$ is the dark matter energy density, while $p_r(r)$ and $p_t(r)$ denote the radial and tangential pressures, respectively. These quantities are determined self-consistently from the Einstein field equations.

For the metric in Eq.~\eqref{1stmetric}, the independent components of the Einstein field equations are
\begin{itemize}
    \item \textbf{Temporal component:}
    \begin{equation}
    G_{tt}
    =
    -\frac{h(r)}{r^2}
    \left[rh'(r)+h(r)-1\right]
    =
    8\pi h(r)\rho_{\rm DM}(r),
    \end{equation}

    \item \textbf{Radial component:}
    \begin{equation}
    G_{rr}
    =
    \frac{1}{r^2h(r)}
    \left[rh'(r)+h(r)-1\right]
    =
    8\pi\frac{p_r(r)}{h(r)},
    \end{equation}

    \item \textbf{Angular components:}
    \begin{equation}
    G_{\theta\theta}
    =
    \frac{G_{\phi\phi}}{\sin^2\theta}
    =
    \frac{r^2}{2}h''(r)+\frac{r}{2}h'(r)
    =
    8\pi r^2p_t(r).
    \end{equation}
\end{itemize}

Here, $h'(r)=dh/dr$ and $h''(r)=d^2h/dr^2$. The temporal and radial components have opposite signs on their geometric sides, which immediately leads to
\begin{equation}
p_r(r)=-\rho_{\rm DM}(r).
\end{equation}
Thus, within the present Schwarzschild-like metric ansatz, the dark matter source is necessarily anisotropic and possesses a radial tension.

The temporal component of the Einstein field equations yields the first-order differential equation
\begin{equation}
\frac{d}{dr}\left[r(1-h(r))\right]
=
8\pi r^2\rho_{\rm DM}(r),
\end{equation}
which can be integrated once the dark matter density profile is specified. Imposing the Schwarzschild limit in the absence of dark matter gives
\begin{equation}
h(r)
=
1-\frac{r_s}{r}
-\frac{8\pi}{r}
\int_0^r \rho_{\rm DM}(r')\,r'^2\,dr',
\end{equation}
where $r_s=2M$ is the Schwarzschild radius associated with the central black hole. The integral represents the contribution of the surrounding dark matter halo to the metric function.

For the King density profile given in Eq.~\eqref{rho}, the integral can be evaluated analytically. The resulting static, spherically symmetric black hole spacetime is
\begin{gather}
ds^2
=
-h(r)dt^2+\frac{dr^2}{h(r)}+r^2d\Omega_2^2,
\\
h(r)
=
1-\frac{r_s}{r}
+\frac{8\pi\rho_0r_0^3}{r}
\left[
\ln\left(
\sqrt{1+\left(\frac{r}{r_0}\right)^2}
-\frac{r}{r_0}
\right)
+\frac{r}{\sqrt{r_0^2+r^2}}
\right].
\label{metric}
\end{gather}

The solution presented above is different to the one in Ref.~\cite{Al-Badawi:2026cjx},
\begin{equation}
 h(r)
=\left(\sqrt{\frac{r^2}{r_0^2}+1}+\frac{r^2}{r_0^2}\right) ^{\frac{8\pi\rho_0r_0^3}{r}}-\frac{r_s}{r}.
\end{equation}

The first term in $h(r)$ gives the asymptotically flat contribution, while the Schwarzschild term $-r_s/r$ describes the gravitational field of the central black hole. The remaining terms represent the modification of the spacetime geometry induced by the King dark matter halo. In particular, the dark matter contribution vanishes when $\rho_0\rightarrow0$, and Eq.~\eqref{metric} reduces to the Schwarzschild solution,
\begin{equation}
h(r)\rightarrow1-\frac{r_s}{r}.
\end{equation}

\section{Ray Tracing}

We now investigate the propagation of photons in the spacetime of a Schwarzschild black hole surrounded by a King dark matter halo, described by the metric in Eq.~\eqref{metric}. For a massless particle, the geodesic Lagrangian is
\begin{align}
\mathcal{L}(x^\alpha,\dot{x}^\alpha)
&=
\frac{1}{2}g_{\mu\nu}\dot{x}^\mu\dot{x}^\nu \nonumber\\
&=
\frac{1}{2}
\left[
-h(r)\dot{t}^{\,2}
+\frac{\dot{r}^{\,2}}{h(r)}
+r^2\dot{\theta}^{\,2}
+r^2\sin^2\theta\,\dot{\phi}^{\,2}
\right]
=0,
\label{Lagrangian}
\end{align}
where the vanishing of the Lagrangian follows from the null condition
$g_{\mu\nu}\dot{x}^\mu\dot{x}^\nu=0$.

Since the Lagrangian is independent of the coordinates $t$ and $\phi$, the corresponding canonical momenta are conserved. These constants of motion are identified with the photon energy $E$ and angular momentum $L$,
\begin{gather}
p_t
=
\frac{\partial\mathcal{L}}{\partial\dot{t}}
=
-h(r)\dot{t}
=
-E,
\label{4}\\
p_\phi
=
\frac{\partial\mathcal{L}}{\partial\dot{\phi}}
=
r^2\sin^2\theta\,\dot{\phi}
=
L.
\label{5}
\end{gather}

Owing to the spherical symmetry of the spacetime, the photon trajectory can always be chosen to lie in a plane. Without loss of generality, we take this to be the equatorial plane,
$\theta=\pi/2$, with $\dot{\theta}=0$. The Lagrangian then reduces to
\begin{align}
\mathcal{L}(r,\dot{t},\dot{r},\dot{\phi})
&=
\frac{1}{2}
\left[
-h(r)\dot{t}^{\,2}
+\frac{\dot{r}^{\,2}}{h(r)}
+r^2\dot{\phi}^{\,2}
\right]
\nonumber\\
&=
\frac{1}{2}
\left[
-\frac{E^2}{h(r)}
+\frac{\dot{r}^{\,2}}{h(r)}
+\frac{L^2}{r^2}
\right]
=0.
\label{Lplane}
\end{align}

The radial motion of the photon is therefore governed by
\begin{equation}
\dot{r}^{\,2}
+
\frac{L^2}{r^2}h(r)
=
E^2.
\label{help}
\end{equation}
Introducing the impact parameter
\begin{equation}
b=\frac{L}{E},
\end{equation}
the effective potential can be written as
\begin{align}
\frac{V_{\rm eff}}{E^2}
&=
\frac{b^2h(r)}{r^2}
\nonumber\\
&=
\frac{b^2}{r^2}
\left\{
1-\frac{r_s}{r}
+\frac{8\pi\rho_0r_0^3}{r}
\left[
\ln\left(
\sqrt{1+\left(\frac{r}{r_0}\right)^2}
-\frac{r}{r_0}
\right)
+\frac{r}{\sqrt{r_0^2+r^2}}
\right]
\right\}.
\end{align}

The effective potential determines the radial motion of photons and provides the basis for studying circular photon orbits, the photon sphere, gravitational lensing, and black-hole shadows. In particular, an unstable circular photon orbit is located at an extremum of the effective potential.

For a circular photon orbit at $r=r_c$, the radial velocity and radial acceleration must vanish. Equivalently, the two conditions
\begin{equation}
\left.\dot{r}\right|_{r=r_c}=0,
\qquad
\left.\frac{dV_{\rm eff}}{dr}\right|_{r=r_c}=0
\end{equation}
give
\begin{equation}
\frac{V_{\rm eff}(r_c)}{E^2}=1,
\label{new2}
\end{equation}
and
\begin{equation}
V_{\rm eff}'(r_c)=0.
\end{equation}
Using
\begin{equation}
\mathrm{arcsinh}(z)
=
\ln\left(z+\sqrt{1+z^2}\right),
\end{equation}
the metric function may equivalently be expressed as
\begin{equation}
h(r)
=
1-\frac{r_s}{r}
+\frac{8\pi\rho_0r_0^3}{r}
\left[
\frac{r}{\sqrt{r_0^2+r^2}}
-\mathrm{arcsinh}\left(\frac{r}{r_0}\right)
\right].
\end{equation}
Consequently, the condition \eqref{new2} becomes
\begin{equation}
\frac{b^2}{r_c^2}
\left[
1-\frac{r_s}{r_c}
+\frac{8\pi\rho_0r_0^3}{r_c}
\left(
\frac{r_c}{\sqrt{r_0^2+r_c^2}}
-\mathrm{arcsinh}\left(\frac{r_c}{r_0}\right)
\right)
\right]
=1.
\label{new22}
\end{equation}

The extremum condition can be written explicitly as
\begin{multline}
\frac{V_{\rm eff}'(r)}{E^2}
=
\frac{b^2}{r^4}
\Biggl[
3r_s-2r
-\frac{8\pi r^3r_0^3\rho_0}{(r^2+r_0^2)^{3/2}}
-\frac{24\pi r_0^3\rho_0r}{\sqrt{r^2+r_0^2}}
\\
\hspace{3cm}
+24\pi r_0^3\rho_0
\,\mathrm{arcsinh}\left(\frac{r}{r_0}\right)
\Biggr]
=0.
\label{new1}
\end{multline}

Combining Eqs.~\eqref{new22} and \eqref{new1} eliminates both the Schwarzschild-radius contribution and the $\mathrm{arcsinh}$ term, yielding the exact relation
\begin{equation}
\left(r_c^2+r_0^2\right)^{3/2}
\left(
1-\frac{3r_c^2}{b^2}
\right)
-
8\pi\rho_0r_0^3r_c^2
=0.
\label{quartic1}
\end{equation}
This equation, together with Eq.~\eqref{new2}, determines the circular photon-orbit radius $r_c$ and the corresponding critical impact parameter $b_c$. In particular, the photon-sphere radius is not independently shifted by varying $b$; rather, $b$ takes the critical value associated with the circular orbit.

For a weak dark matter contribution and in the regime $r_c\gg r_0$, Eq.~\eqref{quartic1} can be expanded as
\begin{equation}
\left(r_c^2+r_0^2\right)^{3/2}
\approx r_c^3
\left[
1+\frac{3r_0^2}{2r_c^2}
+\mathcal{O}\left(\frac{r_0^4}{r_c^4}\right)
\right].
\end{equation}
Retaining the leading term gives
\begin{equation}
q(r_c)
\equiv
r_c\left(1-\frac{3r_c^2}{b^2}\right)
-8\pi\rho_0r_0^3
=0,
\label{quartic1_approx}
\end{equation}
where we have divided Eq.~\eqref{quartic1} by $r_c^2$. The solution $r_c$ gives the radius of the corresponding circular photon orbit within this leading-order approximation.

\begin{figure}[h]
    \centering
    \includegraphics[scale=0.9]{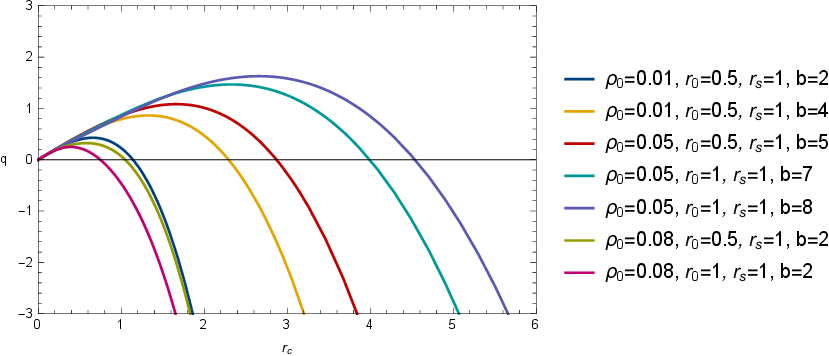}
    \caption{Profile of $q(r_c)$ for selected dark matter parameters and impact parameters.}
    \label{2}
\end{figure}

Figure~\ref{2} illustrates the behavior of the function $q(r_c)$ for representative choices of the dark matter and photon parameters. The positive root of this equation provides the corresponding circular-orbit radius within the adopted approximation. The exact circular-orbit condition, however, is given by Eq.~\eqref{quartic1} and should be used when the characteristic halo scale $r_0$ is not negligible compared with $r_c$.

We next consider the spatial trajectories of photons. From Eq.~\eqref{help}, together with
\begin{equation}
\dot{r}
=
\frac{dr}{d\phi}\dot{\phi},
\end{equation}
and
\begin{equation}
\frac{1}{b}
=
\frac{E}{L}
=
\frac{h(r)}{r^2\,d\phi/dt},
\end{equation}
we obtain
\begin{align}
\left(\frac{dr}{d\phi}\right)^2
&=
\frac{r^4}{L^2}
\left(
E^2-\frac{L^2}{r^2}h(r)
\right)
\nonumber\\
&=
\frac{r^4}{b^2}-r^2h(r)
\nonumber\\
&=
r^2h(r)
\left(
\frac{r^2}{b^2h(r)}-1
\right).
\end{align}

\begin{figure}[h]
    \centering
    \includegraphics[scale=0.9]{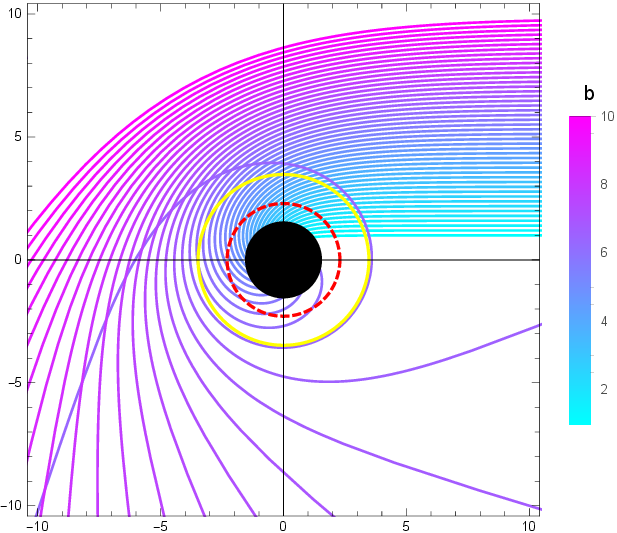}
    \caption{Ray-tracing profile of a black hole--King dark matter system with $r_s=2$, $\rho_0=0.0179$, and $r_0=1$.}
    \label{lt}
\end{figure}

Figure~\ref{lt} displays representative photon trajectories for impact parameters in the range $1\leq b\leq10$ for the black hole--dark matter configuration specified by
$r_s=2$, $\rho_0=0.0179$, and $r_0=1$. The central black hole is indicated by the solid black circle, while the red dashed curve denotes the event horizon at $r=2.293$. As the impact parameter increases, the gravitational deflection decreases and the photon trajectories become progressively less curved. At the critical impact parameter $b=b_c$, separating captured and scattered trajectories, the photon asymptotically approaches the unstable circular photon orbit, represented by the yellow circle at $r=3.473$. For $b<b_c$, the photon is captured by the black hole, whereas for $b>b_c$, it is scattered back to infinity.

\subsection{Black Hole Shadow}

The photon sphere is defined by the existence of an unstable circular null orbit. For the radial equation of motion,
\begin{equation}
\left(\frac{dr}{d\phi}\right)^2
=
\frac{r^4}{b^2}-r^2h(r),
\end{equation}
a circular photon orbit at $r=r_{ps}$ satisfies
\begin{equation}
\left.\frac{dr}{d\phi}\right|_{r=r_{ps}}=0,
\qquad
\left.\frac{d^2r}{d\phi^2}\right|_{r=r_{ps}}=0.
\end{equation}
The first condition gives
\begin{equation}
\frac{r_{ps}^2}{b_c^2 h(r_{ps})}=1,
\label{psshort}
\end{equation}
while the second condition yields
\begin{equation}
\left.\frac{d}{dr}\left(\frac{r^2}{h(r)}\right)\right|_{r=r_{ps}}=0.
\label{ps}
\end{equation}
The latter can equivalently be written as
\begin{equation}
2h(r_{ps})-r_{ps}h'(r_{ps})=0.
\label{ps_condition}
\end{equation}
This condition is equivalent to requiring that the effective potential admit an extremum at the photon-sphere radius.

Substituting the King-halo metric function \eqref{metric} into Eq.~\eqref{ps_condition}, we obtain
\begin{equation}
-3r_s+2r
+8\pi\rho_0r_0^3
\left[
\frac{r(3r^2+2r_0^2)}{(r^2+r_0^2)^{3/2}}
+\frac{r}{\sqrt{r^2+r_0^2}}
-3\,\mathrm{arcsinh}\left(\frac{r}{r_0}\right)
\right]
=0.
\label{photonsphereeq}
\end{equation}
In the absence of dark matter, $\rho_0=0$, Eq.~\eqref{photonsphereeq} reduces to
\begin{equation}
-3r_s+2r=0,
\end{equation}
and hence
\begin{equation}
r_{ps}=\frac{3}{2}r_s.
\end{equation}
For a Schwarzschild black hole, this corresponds to the familiar photon-sphere radius $r_{ps}=3M$.

The critical impact parameter follows directly from Eq.~\eqref{psshort},
\begin{equation}
b_c
=
\frac{r_{ps}}{\sqrt{h(r_{ps})}},
\label{bc}
\end{equation}
where $r_{ps}$ is determined by Eq.~\eqref{photonsphereeq}. The critical impact parameter separates photon trajectories that are captured by the black hole from those that are scattered back to infinity. In particular, $b<b_c$ corresponds to photon capture, whereas $b>b_c$ corresponds to scattering.

For an observer at spatial infinity, the angular radius of the shadow is determined by the critical impact parameter. Since $h(r)\rightarrow1$ as $r\rightarrow\infty$, the shadow radius is
\begin{equation}
r_{ps}
=
b_c\sqrt{h(\infty)}
=
\frac{r_{ps}}{\sqrt{h(r_{ps})}}.
\label{shadowradius}
\end{equation}
Thus, the shadow size is determined entirely by the metric function evaluated at the photon sphere.

For a weak dark matter contribution, the photon-sphere radius can be obtained perturbatively by writing
\begin{equation}
r_{ps}
=
\frac{3}{2}r_s+\delta r,
\qquad
\delta r=\mathcal{O}(\rho_0r_0^3).
\end{equation}
Expanding Eq.~\eqref{photonsphereeq} to first order in $\rho_0r_0^3$ gives
\begin{equation}
r_{ps} \approx \frac{1}{2} \left( 3 r_s - 32 \pi r_0^3 \rho_0 \right).
\label{rps_approx}
\end{equation}

The corresponding shadow radius can be expanded as
\begin{equation}
r_{ps}
\approx
\frac{3}{2} \sqrt{3} \Biggl[ r_s + 8 \pi r_0^3 \rho_0 \Bigl( -1 + \ln \frac{3 r_s}{r_0} \Bigr) \Biggr].
\end{equation}
where $\delta r_{ps}$ is obtained by consistently expanding Eq.~\eqref{shadowradius} to first order in $\rho_0r_0^3$. In the absence of dark matter, one immediately recovers
\begin{equation}
r_{ps}
=
\frac{3\sqrt{3}}{2}r_s,
\end{equation}
which is the standard Schwarzschild shadow radius \cite{Perlick:2021aok}.

\begin{figure}[H]
    \centering
    \includegraphics[scale=0.52]{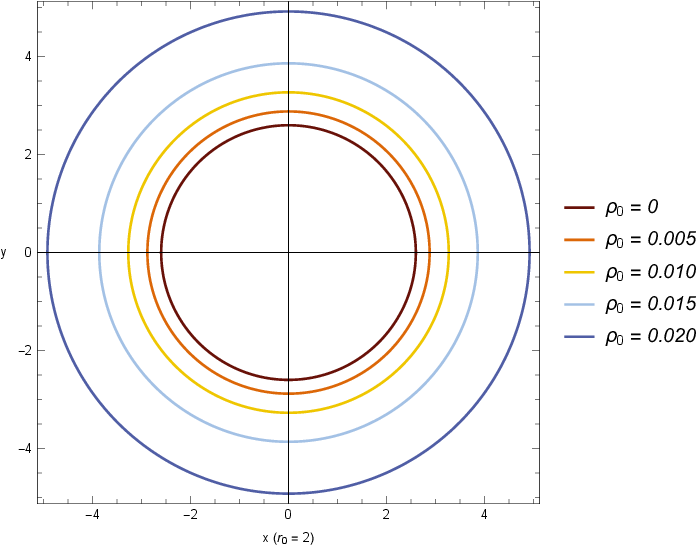}
    \includegraphics[scale=0.52]{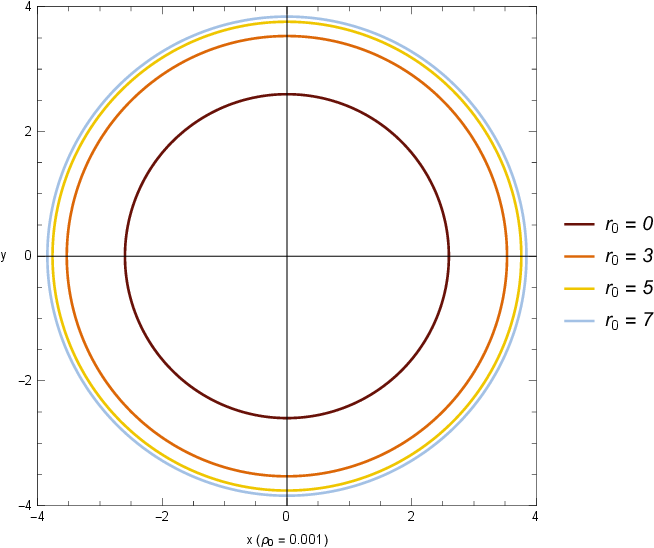}\\
    \caption{Profiles of the shadow radius $r_{ps}$ for different combinations of $\{\rho_0,r_0\}$ with $r_s=1$.}
    \label{R}
\end{figure}

Figure~\ref{R} illustrates the variation of the shadow radius $r_{ps}$ for different values of the King-halo parameters $\rho_0$ and $r_0$, with the Schwarzschild radius fixed at $r_s=1$. The left panel shows the dependence on the central density $\rho_0$ for fixed $r_0=2$, while the right panel displays the dependence on the scale radius $r_0$ for fixed $\rho_0=0.001$. These variations demonstrate how the dark matter halo modifies both the photon-sphere radius and the critical impact parameter, thereby changing the apparent size of the black hole shadow.

\subsection{Weak Deflection Angle}

We now investigate the weak gravitational deflection of light in the static black hole spacetime embedded in a King dark matter halo. We employ the Gauss--Bonnet theorem applied to the optical metric, following the approach introduced by Gibbons and Werner \cite{Gibbons:2008rj,Waseem:2025yib}. This method relates the deflection angle to the Gaussian curvature of the optical geometry.

For null geodesics in the equatorial plane, $ds^2=0$ gives the two-dimensional optical metric
\begin{equation}
dt^2
=
\frac{dr^2}{h^2(r)}
+
\frac{r^2d\phi^2}{h(r)}.
\label{opticalmetric}
\end{equation}
The determinant of the optical metric is therefore
\begin{equation}
g_{\rm opt}
=
\frac{r^2}{h^3(r)}.
\end{equation}
The Gaussian curvature of this two-dimensional optical geometry is
\begin{equation}
K
=
\frac{1}{2}
\left[
\frac{1}{2}\left(\frac{dh}{dr}\right)^2
-h(r)\frac{d^2h}{dr^2}
\right].
\end{equation}

In the weak-field regime, the leading-order Gauss--Bonnet expression for the deflection angle is
\begin{equation}
\alpha
=
-\int_0^\pi
\int_{b/\sin\phi}^{\infty}
K\sqrt{g_{\rm opt}}\,
dr\,d\phi.
\label{GB}
\end{equation}
The minus sign is required because the Gaussian curvature of the Schwarzschild optical geometry is negative, while the physical deflection angle is positive.

Substituting the King-halo metric function into $K\sqrt{g_{\rm opt}}$ and expanding in the weak-field regime gives the corresponding perturbative expression,
\begin{equation} K \sqrt{g_{opt}} \approx \frac{r_s}{r^2} + \frac{2 \pi r_0^3 \rho_0}{r^2} \Biggl[ 4 \ln \frac{2 r}{r_0} + \frac{6 r_s}{r} \ln \frac{2 r}{r_0}\Biggr]. 
\end{equation}

The resulting deflection angle can then be written schematically as
\begin{equation}
\alpha \approx \frac{1}{b} \Biggl[ 2 r_s + 8 \pi r_0^3 \rho_0 \Bigl( -1 + 2 \ln \frac{b}{r_0} \Bigr) \Biggr].
\end{equation}
where $\alpha_{\rm DM}$ denotes the leading correction induced by the King dark matter halo.

In the absence of dark matter, $\rho_0=0$, the Gaussian curvature reduces to the Schwarzschild result and Eq.~\eqref{GB} gives
\begin{equation}
\alpha
=
\frac{2r_s}{b},
\end{equation}
as expected \cite{Waseem:2025yib}.

\begin{figure}[H]
    \centering
    \includegraphics[scale=0.9]{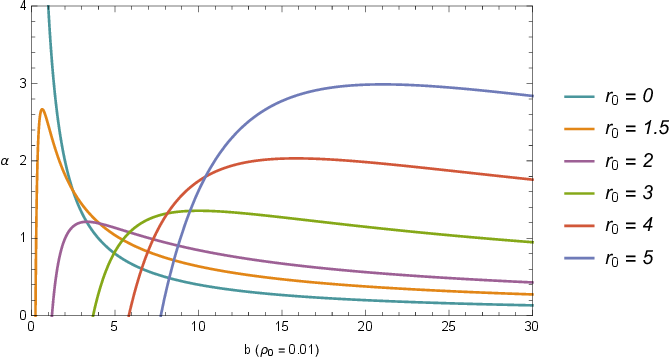}
    \caption{Variation of the deflection angle $\alpha$ for different values of $\{ \rho_0=0.01,r_0\}$ with $r_s=1$.}
    \label{Lensing}
\end{figure}

Figure~\ref{Lensing} shows the variation of the deflection angle $\alpha$ with the impact parameter $b$ for black holes embedded in a King dark matter halo, with $r_s=1$ and $\rho_0=0.01$, for different values of the halo scale radius $r_0$.

\section{Orbit Stability and Lyapunov Exponent}

As discussed in the previous section, a circular photon orbit corresponds to an extremum of the effective potential. For the photon sphere, this extremum is a maximum, indicating that the orbit is unstable against small radial perturbations. The instability can be quantified by the Lyapunov exponent, which characterizes the exponential growth rate of these perturbations. In the eikonal limit, the Lyapunov exponent defined with respect to the coordinate time is also related to the imaginary part of the quasinormal-mode frequencies.

The photon dynamics can be described by the Hamiltonian \eqref{Lplane},
\begin{align}
H=\mathcal{L}
&=\frac{1}{2}\left[
-\frac{E^2}{h(r)}
+h(r)p_r^2
+\frac{V_{\rm eff}(r)}{h(r)}
\right] \nonumber\\
&=\frac{1}{2}g^{\mu\nu}p_\mu p_\nu ,
\end{align}
where $E$ is the conserved photon energy, $p_r$ is the radial canonical momentum, and $V_{\rm eff}(r)$ denotes the effective potential associated with the angular motion.

The Hamilton equations in the equatorial plane are
\begin{gather}
p_r=\frac{\partial\mathcal{L}}{\partial\dot r}
=\frac{\dot r}{h(r)},
\qquad
\dot r=\frac{\partial H}{\partial p_r}
=h(r)p_r ,
\\
\dot p_r
=-\frac{\partial H}{\partial r}
=-\frac{1}{2}
\left[
h'(r)p_r^2
+\frac{V_{\rm eff}'(r)}{h(r)}
-\frac{h'(r)}{h^2(r)}
\left(-E^2+V_{\rm eff}(r)\right)
\right].
\end{gather}

For a circular photon orbit at $r=r_c$, the radial motion satisfies
\begin{equation}
\dot r=0,
\qquad
V_{\rm eff}'(r_c)=0,
\qquad
V_{\rm eff}(r_c)=E^2.
\end{equation}
Consequently, $p_r=0$ on the circular orbit. To investigate its stability, we introduce small radial perturbations,
\begin{equation}
r=r_c+\delta r,
\qquad
p_r=\delta p_r,
\end{equation}
and retain only terms linear in the perturbations. The resulting linearized equations are
\begin{gather}
\delta\dot r=h(r_c)\,\delta p_r,
\\
\delta\dot p_r
=-\frac{V_{\rm eff}''(r_c)}{2h(r_c)}\,\delta r.
\end{gather}
Here, the terms proportional to $p_r^2$ are second order in the perturbations and therefore do not contribute at linear order.

The linearized system can be expressed in matrix form as
\begin{equation}
\frac{d}{d\lambda}
\begin{pmatrix}
\delta r\\
\delta p_r
\end{pmatrix}
=
\begin{pmatrix}
0 & h(r_c)\\
-\dfrac{V_{\rm eff}''(r_c)}{2h(r_c)} & 0
\end{pmatrix}
\begin{pmatrix}
\delta r\\
\delta p_r
\end{pmatrix},
\end{equation}
where $\lambda$ is the affine parameter along the photon trajectory. The eigenvalues of the stability matrix are determined by
\begin{equation}
\det
\left[
\begin{pmatrix}
0 & h(r_c)\\
-\dfrac{V_{\rm eff}''(r_c)}{2h(r_c)} & 0
\end{pmatrix}
-\Lambda I
\right]=0,
\end{equation}
which gives
\begin{equation}
\Lambda^2
=-\frac{V_{\rm eff}''(r_c)}{2}.
\end{equation}

For the effective potential
\begin{equation}
V_{\rm eff}(r)=L^2\frac{h(r)}{r^2},
\end{equation}
the Lyapunov exponent becomes
\begin{align}
\Lambda^2
&=-\frac{L^2}{2}
\left.
\frac{d^2}{dr^2}
\left(\frac{h(r)}{r^2}\right)
\right|_{r=r_c}
\nonumber\\
&=
L^2\left[
-\frac{h''(r_c)}{2r_c^2}
+\frac{2h'(r_c)}{r_c^3}
-\frac{3h(r_c)}{r_c^4}
\right].
\end{align}

The sign of $\Lambda^2$ determines the stability of the circular photon orbit. For
\begin{equation}
V_{\rm eff}''(r_c)>0,
\end{equation}
the orbit is stable and $\Lambda^2<0$, corresponding to oscillatory perturbations. Conversely, for
\begin{equation}
V_{\rm eff}''(r_c)<0,
\end{equation}
the orbit is unstable and $\Lambda^2>0$, so that radial perturbations grow exponentially with the affine parameter. Thus, the curvature of the effective potential at the circular orbit determines the stability and the characteristic instability rate of the photon trajectory.

\begin{figure}[h]
    \centering
    \includegraphics[scale=0.9]{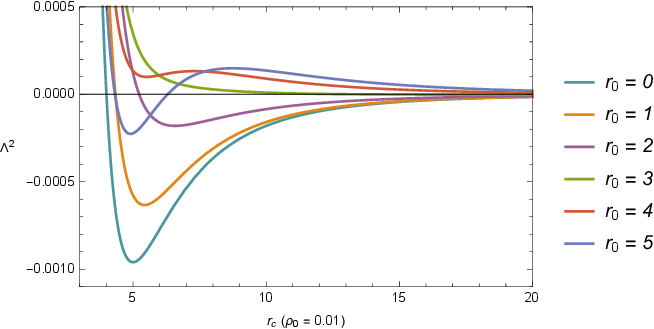}
    \caption{Profile of $\Lambda^2$ as a function of the circular photon-orbit radius for different values of the dark-matter halo scale radius $r_0$, with $\rho_0=0.01$.}
    \label{LE}
\end{figure}

Figure~\ref{LE} shows $\Lambda^2$ as a function of the circular-orbit radius for different values of the halo scale radius $r_0$, with the dark-matter density fixed at $\rho_0=0.01$. The variation of $\Lambda^2$ with $r_0$ demonstrates that the dark-matter halo profile modifies the instability rate of circular photon orbits and hence their characteristic instability timescale.

\section{Scalar Quasinormal Modes}

Quasinormal modes (QNMs) describe the damped oscillations of perturbations around black holes and other compact objects. They are characterized by complex frequencies, with the real part determining the oscillation frequency and the imaginary part determining the damping rate. The QNM spectrum therefore provides important information about the stability and dynamical response of the spacetime to perturbations.

In this work, we consider scalar perturbations of a spherically symmetric black hole immersed in a King dark matter halo. The dynamics of the scalar field are governed by the covariant Klein--Gordon equation,
\begin{equation}
\left[
\frac{1}{\sqrt{-g}}
\partial_\mu
\left(
\sqrt{-g}\,g^{\mu\nu}\partial_\nu
\right)
-m^2
\right]\psi=0,
\end{equation}
where $m$ is the scalar-field mass and $g$ is the determinant of the metric tensor.

Using the metric \eqref{metric} and exploiting spherical symmetry, we adopt the separable ansatz
\begin{equation}
\psi(t,r,\theta,\phi)
=e^{-i\omega t}R(r)Y_l^{m_l}(\theta,\phi),
\end{equation}
where $Y_l^{m_l}(\theta,\phi)$ are the spherical harmonics. Substitution into the Klein--Gordon equation gives the radial equation
\begin{equation}
\partial_r
\left[
h(r)r^2\partial_r R(r)
\right]
+
\left[
\frac{\omega^2r^2}{h(r)}
-l(l+1)
-m^2r^2
\right]R(r)=0,
\label{radial}
\end{equation}
which determines the radial behavior of the scalar perturbation.

We now consider a massless scalar field, $m=0$, and introduce the tortoise coordinate $r^*$ through
\begin{equation}
\frac{dr^*}{dr}=\frac{1}{h(r)},
\end{equation}
together with the field redefinition
\begin{equation}
R(r)=\frac{1}{r}\mathcal{R}(r^*).
\end{equation}
Equation~\eqref{radial} then reduces to the Schr\"odinger-like equation
\begin{equation}
\frac{d^2\mathcal{R}}{dr^{*2}}
+
\left[
\omega^2-V(r)
\right]\mathcal{R}=0,
\end{equation}
where the effective potential is
\begin{equation}
V(r)
=
\frac{h(r)}{r^2}
\left[
l(l+1)+rh'(r)
\right].
\end{equation}

In the eikonal limit, $l\gg1$, the leading contribution to the effective potential is
\begin{equation}
V(r)\approx 
\frac{h(r)}{r^2}l^2.
\end{equation}
Since the photon effective potential introduced previously is
\begin{equation}
V_{\rm eff}(r)
=
L^2\frac{h(r)}{r^2},
\end{equation}
the radial equation can be written, to leading order in $l$, as
\begin{gather}
\frac{d^2\mathcal{R}}{dr^{*2}}
+W(r)\mathcal{R}=0,
\\
W(r)
=
\omega^2-\frac{h(r)}{r^2}l^2
=
\omega^2-\frac{l^2}{L^2}V_{\rm eff}(r).
\end{gather}
Thus, in the eikonal regime, the scalar QNM spectrum is directly governed by the same effective potential that determines the circular null geodesics.

The QNM frequencies can be estimated using the leading-order WKB approximation. Following the standard Schutz--Will/Iyer--Will approach \cite{Schutz:1985km}, the quantization condition is
\begin{equation}
\frac{W(r_0)}
{\sqrt{2W^{(2)}(r_0)}}
=
-i\left(n+\frac{1}{2}\right),
\label{WKB1}
\end{equation}
where $n=0,1,2,\ldots$ is the overtone number and
\begin{equation}
W^{(2)}(r_0)
=
\left.
\frac{d^2W}{d{r^*}^2}
\right|_{r=r_0}.
\end{equation}
Here, $r_0$ denotes the location of the maximum of the effective potential. In the eikonal limit, this point coincides with the radius of the unstable circular null orbit,
\begin{equation}
r_0=r_c.
\end{equation}
Equivalently, $W(r)$ has a minimum at $r=r_c$, since $W=\omega^2-V$.

Solving the WKB condition to leading order gives
\begin{equation}
\omega_{\rm QNM}
\approx 
\frac{l}{|L|}
\sqrt{V_{\rm eff}(r_c)}
-i\left(n+\frac{1}{2}\right)
\sqrt{
-\frac{1}{2V_{\rm eff}(r_c)}
\left.
\frac{d^2V_{\rm eff}}{d{r^*}^2}
\right|_{r=r_c}
}.
\end{equation}
At the circular null orbit, $V_{\rm eff}'(r_c)=0$, and hence
\begin{equation}
\left.
\frac{d^2V_{\rm eff}}{d{r^*}^2}
\right|_{r=r_c}
=
\left.
h(r)\frac{d}{dr}
\left[
h(r)\frac{dV_{\rm eff}}{dr}
\right]
\right|_{r=r_c}
=
h^2(r_c)V_{\rm eff}''(r_c).
\end{equation}

The above expression establishes the correspondence between the eikonal QNM spectrum and the properties of the unstable photon orbit. The real part of the QNM frequency is determined by the value of the effective potential at the circular null orbit, whereas the imaginary part is governed by its curvature there. Using the Lyapunov exponent $\Lambda$ introduced in the previous section, defined with respect to the affine parameter $\lambda$ by
\begin{equation}
\Lambda^2
=
-\frac{1}{2}V_{\rm eff}''(r_c),
\end{equation}
the coordinate-time instability exponent is
\begin{equation}
\lambda_t
=
\frac{h(r_c)}{E}\,|\Lambda|,
\end{equation}
where $E=\sqrt{V_{\rm eff}(r_c)}$ for the circular null orbit. Therefore,
\begin{equation}
\lambda_t
=
\frac{|\Lambda|}{L^2}
r_c^2\sqrt{V_{\rm eff}(r_c)}.
\end{equation}
The eikonal QNM frequency consequently takes the familiar form
\begin{equation}
\omega_{\rm QNM}
=
\frac{l}{|L|}
\sqrt{V_{\rm eff}(r_c)}
-i\left(n+\frac{1}{2}\right)\lambda_t,
\end{equation}
or, equivalently,
\begin{equation}
\omega_{\rm QNM}
=
\frac{l}{|L|}
\sqrt{V_{\rm eff}(r_c)}
-i\left(n+\frac{1}{2}\right)
\frac{|\Lambda|}{L^2}
r_c^2\sqrt{V_{\rm eff}(r_c)}.
\end{equation}

For the regime $\rho_0r_0^2\ll1$, the QNM frequency can be expanded perturbatively in the dark-matter contribution. To leading order, we obtain
\begin{multline}
\omega_{\rm QNM}
=
\frac{2l}{3\sqrt{3}\,r_s^2}
\left[
r_s
+
8\pi r_0^3\rho_0
\left(
1-\ln\frac{3r_s}{r_0}
\right)
\right]
\\
-i\frac{2n+1}{3\sqrt{3}\,r_s^2}
\left[
r_s
+
8\pi r_0^3\rho_0
\left(
1-\ln\frac{3r_s}{r_0}
\right)
\right].
\label{QNMApprox}
\end{multline}

As a consistency check, consider the Schwarzschild limit, for which
\begin{equation}
h(r)=1-\frac{2M}{r}
\end{equation}
and
\begin{equation}
V_{\rm eff}(r)
=
\frac{L^2}{r^2}
\left(
1-\frac{2M}{r}
\right).
\end{equation}
The circular null orbit is located at
\begin{equation}
r_c=3M.
\end{equation}
At this radius,
\begin{equation}
\Lambda=\frac{|L|}{9\sqrt{3}\,M^2},
\qquad
V_{\rm eff}(r_c)
=
\frac{L^2}{27M^2}.
\end{equation}
Consequently, the general eikonal expression reduces to the standard Schwarzschild result \cite{Ferrari:1984zz},
\begin{equation}
\omega_{\rm QNM}^{\rm Schwarzschild}
=
\frac{l}{3\sqrt{3}\,M}
-i\frac{n+\frac{1}{2}}{3\sqrt{3}\,M}.
\end{equation}

The presence of the dark-matter halo therefore produces a leading-order correction to the QNM spectrum. From Eq.~\eqref{QNMApprox}, this correction can be expressed as
\begin{equation}
\Delta\omega_{\rm QNM}
=
\frac{
8\pi r_0^3\rho_0
\left(
1-\ln\frac{3r_s}{r_0}
\right)
}{r_s}
\,
\omega_{\rm QNM}^{\rm Schwarzschild},
\end{equation}
indicating that the leading dark-matter correction is controlled by the dimensionless combination $r_0^3\rho_0/r_s$.

\section{Thermodynamics}

In this section, we investigate the thermodynamic properties of a static, spherically symmetric black hole immersed in a King dark matter halo. The event horizon radius $r_H$ is determined by the condition
\begin{equation}
h(r_H)=0,
\end{equation}
or equivalently $g_{tt}(r_H)=0$. Solving this condition for the black hole mass gives
\begin{equation}
M =
\frac{1}{2}
\Biggl[
r_H
+
8\pi r_0^3\rho_0
\Biggl(
\frac{r_H}{\sqrt{r_0^2+r_H^2}}
-
\operatorname{arcsinh}\left(\frac{r_H}{r_0}\right)
\Biggr)
\Biggr].
\label{mass}
\end{equation}

In the extended black hole thermodynamic framework, the black hole mass is interpreted as the thermodynamic enthalpy \cite{Kastor:2009wy,Kastor:2018cqc}. For the present asymptotically flat configuration, this quantity reduces to the black hole mass in the absence of an additional thermodynamic pressure. The Hawking temperature is determined by the surface gravity at the event horizon,
\begin{align}
T
&=
\frac{h'(r_H)}{4\pi}
\nonumber\\
&=
\frac{1}{4\pi r_H}
-
\frac{2r_0^3\rho_0}{r_H^2}
\Biggl[
\frac{r_H^3}{(r_0^2+r_H^2)^{3/2}}
+
\operatorname{arcsinh}\left(\frac{r_H}{r_0}\right)
+
\ln
\left(
\frac{\sqrt{r_0^2+r_H^2}-r_H}{r_0}
\right)
\Biggr].
\end{align}
The logarithmic and inverse-hyperbolic terms in the square brackets cancel identically according to
\begin{equation}
\operatorname{arcsinh}(x)
+
\ln\left(\sqrt{1+x^2}-x\right)=0,
\end{equation}
so that the temperature can equivalently be written in the simpler form
\begin{equation}
T
=
\frac{1}{4\pi r_H}
-
\frac{2r_0^3\rho_0 r_H}
{(r_0^2+r_H^2)^{3/2}}.
\label{temperature}
\end{equation}

\begin{figure}[H]
    \centering
    \includegraphics[scale=0.9]{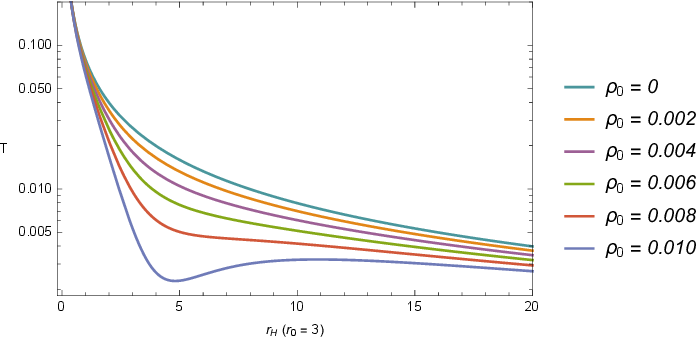}\\
    \includegraphics[scale=0.9]{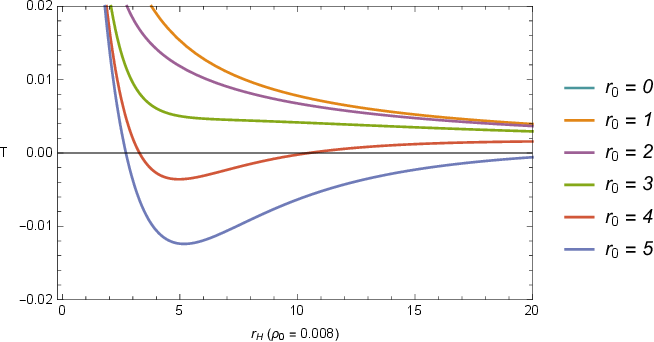}\\
    \caption{Hawking temperature $T$ as a function of the horizon radius $r_H$ for various combinations of the dark matter parameters $\{\rho_0,r_0\}$.}
    \label{T}
\end{figure}

Figure~\ref{T} shows the Hawking temperature $T$ as a function of the horizon radius $r_H$ for different values of the dark matter parameters $\{\rho_0,r_0\}$. In the upper panel, $r_0=3$ is fixed while $\rho_0$ is varied, whereas in the lower panel, $\rho_0=0.008$ is fixed and $r_0$ is varied.

The black hole entropy follows from the first law of thermodynamics,
\begin{align}
S
&=
\int
\frac{1}{T}
\frac{dM}{dr_H}\,dr_H
\nonumber\\
&=
\pi r_H^2
=
\frac{A}{4},
\end{align}
in agreement with the Bekenstein--Hawking area law \cite{Gohain:2024eer}. Thus, the entropy is positive and smooth for all $r_H>0$.

To investigate the local thermodynamic stability of the black hole, we consider its heat capacity at fixed thermodynamic parameters,
\begin{align}
C_H
&=
\frac{\partial M}{\partial T}
=
\frac{\displaystyle \frac{\partial M}{\partial r_H}}
{\displaystyle \frac{\partial T}{\partial r_H}}
\nonumber\\
&=
\frac{
1-\dfrac{8\pi r_0^3r_H^2\rho_0}
{(r_0^2+r_H^2)^{3/2}}
}{
-\Biggl[
\dfrac{1}{2\pi r_H^2}
+
\dfrac{4r_0^3(r_0^2-2r_H^2)\rho_0}
{(r_0^2+r_H^2)^{5/2}}
-
\dfrac{8r_0^3\rho_0}{r_H^3}
\Biggl(
\operatorname{arcsinh}\left(\frac{r_H}{r_0}\right)
+
\ln
\left(
\frac{\sqrt{r_0^2+r_H^2}-r_H}{r_0}
\right)
\Biggr)
\Biggr]
}.
\end{align}
Using the identity above, the heat capacity can also be expressed more compactly as
\begin{equation}
C_H
=
\frac{
1-\dfrac{8\pi r_0^3r_H^2\rho_0}
{(r_0^2+r_H^2)^{3/2}}
}{
-\dfrac{1}{2\pi r_H^2}
+
\dfrac{4r_0^3(2r_H^2-r_0^2)\rho_0}
{(r_0^2+r_H^2)^{5/2}}
}.
\label{heatcapacity}
\end{equation}

\begin{figure}[H]
    \centering
    \includegraphics[scale=0.9]{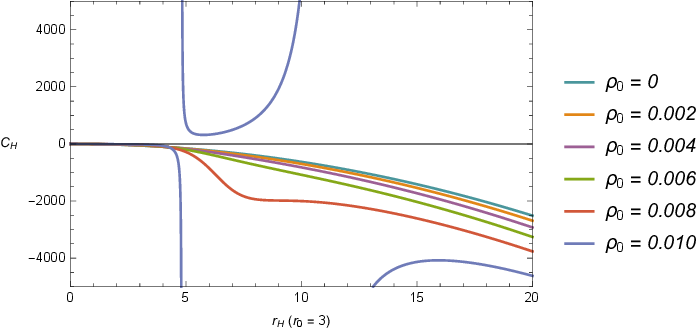}\\
    \includegraphics[scale=0.9]{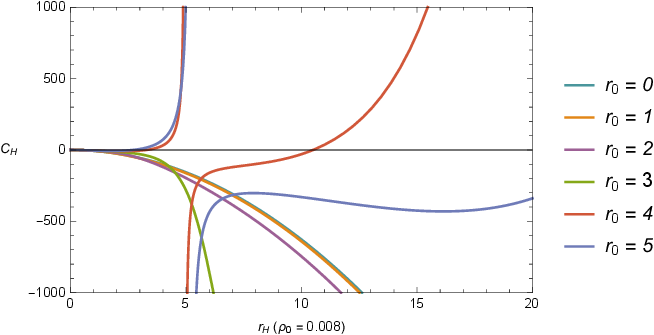}\\
    \caption{Heat capacity $C_H$ as a function of the horizon radius $r_H$ for various combinations of the dark matter parameters $\{\rho_0,r_0\}$.}
    \label{ch}
\end{figure}

A positive heat capacity, $C_H>0$, indicates local thermodynamic stability, whereas $C_H<0$ corresponds to local thermodynamic instability \cite{Mann:2025xrb,Singh:2025svv}. Figure~\ref{ch} displays the heat capacity as a function of $r_H$ for different dark matter halo parameters. In the upper panel, $r_0=3$ is fixed while $\rho_0$ is varied, whereas in the lower panel, $\rho_0=0.008$ is fixed and $r_0$ is varied. In the Schwarzschild limit, $\rho_0=0$, the heat capacity is negative,
\begin{equation}
C_H^{\rm Schwarzschild}=-2\pi r_H^2,
\end{equation}
reflecting the well-known thermodynamic instability of an asymptotically flat Schwarzschild black hole \cite{Benkrane:2025ukw}. For the parameter ranges considered here, the presence of the dark matter halo can generate regions with $C_H>0$, indicating the emergence of locally stable thermodynamic branches. The extent of these stable regions depends on both the central density $\rho_0$ and the halo scale radius $r_0$.

The divergences of $C_H$ occur when
\begin{equation}
\frac{dT}{dr_H}=0,
\end{equation}
corresponding to extrema of the Hawking temperature. Since
\begin{equation}
C_H
=
\frac{dM/dr_H}{dT/dr_H},
\end{equation}
the sign of the heat capacity is determined, provided $dM/dr_H>0$, by the slope of the temperature curve:
\begin{equation}
\frac{dT}{dr_H}>0
\quad\Longrightarrow\quad
C_H>0,
\qquad
\frac{dT}{dr_H}<0
\quad\Longrightarrow\quad
C_H<0.
\end{equation}
Thus, extrema of $T(r_H)$ separate locally stable and unstable thermodynamic branches. For sufficiently small $\rho_0$, if no extremum develops in $T(r_H)$, no locally stable branch is present.

To examine the global thermodynamic behavior of the black hole--dark matter system, we consider the Gibbs free energy,
\begin{align}
G
&=
M-TS
\nonumber\\
&=
\frac{r_H}{4}
+
2\pi r_0^3\rho_0
\Biggl[
\frac{r_H(2r_0^2+3r_H^2)}
{(r_0^2+r_H^2)^{3/2}}
-
\operatorname{arcsinh}\left(\frac{r_H}{r_0}\right)
+
\ln
\left(
\frac{\sqrt{r_0^2+r_H^2}-r_H}{r_0}
\right)
\Biggr].
\end{align}
The sign of $G$ can be used to compare the black hole configuration with the chosen reference background: $G<0$ indicates that the black hole configuration is thermodynamically favored relative to a reference state with vanishing Gibbs free energy, whereas $G>0$ indicates that it is disfavored in this comparison.

\begin{figure}[H]
    \centering
    \includegraphics[scale=0.9]{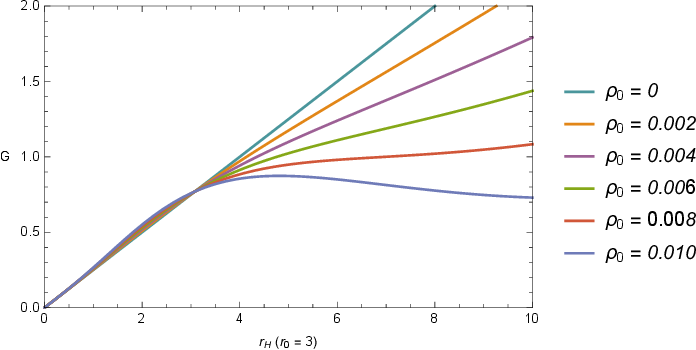}\\
    \includegraphics[scale=0.9]{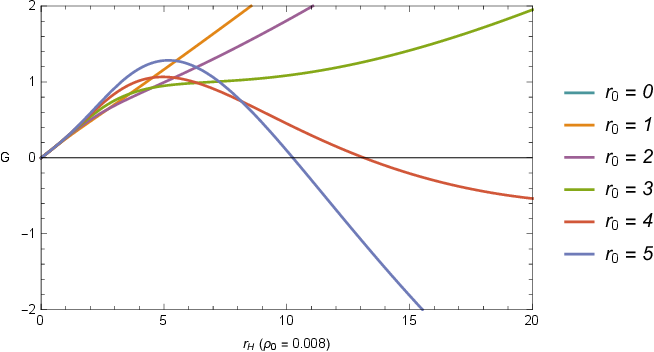}\\
    \caption{Gibbs free energy $G$ as a function of the horizon radius $r_H$ for various combinations of the dark matter parameters $\{\rho_0,r_0\}$. The upper panel shows fixed $r_0=3$ with varying $\rho_0$, while the lower panel shows fixed $\rho_0=0.008$ with varying $r_0$.}
    \label{G}
\end{figure}

Figure~\ref{G} shows the Gibbs free energy as a function of the horizon radius for different combinations of $\{\rho_0,r_0\}$. In the upper panel, $r_0=3$ is fixed while $\rho_0$ is varied, whereas in the lower panel, $\rho_0=0.008$ is fixed and $r_0$ is varied. For the parameter ranges displayed in the figure, increasing the dark matter parameters modifies the regions in which $G$ is positive or negative, indicating that the dark matter halo can significantly affect the global thermodynamic preference of the black hole configuration.

\begin{figure}[H]
    \centering
    \includegraphics[scale=0.9]{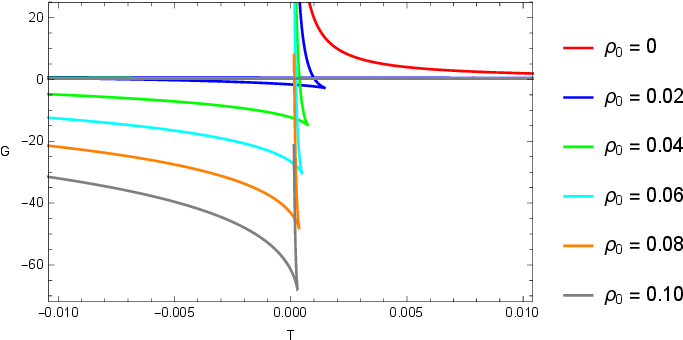}\\
    \includegraphics[scale=0.9]{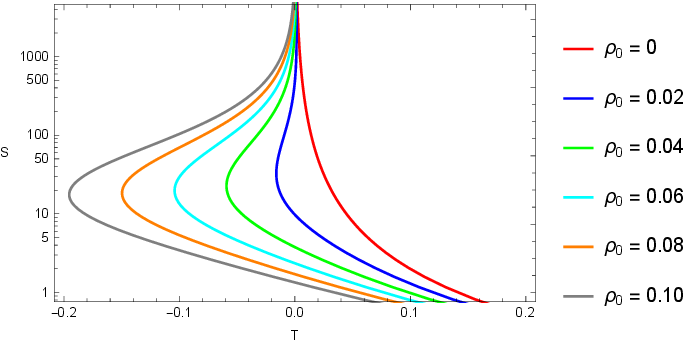}\\
    \includegraphics[scale=0.9]{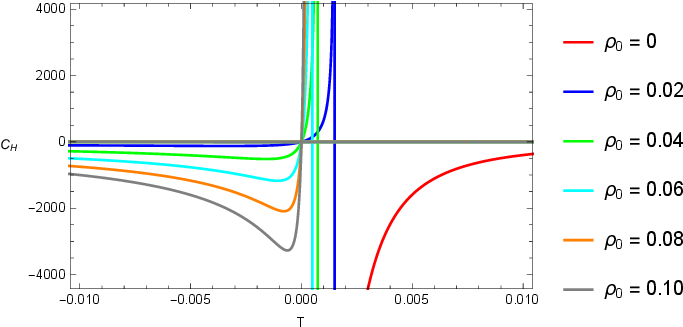}\\
    \caption{Gibbs free energy $G(T)$, entropy $S(T)$, and heat capacity $C_H(T)$ for various values of $\rho_0$ with fixed $r_0=2$.}
    \label{Trans}
\end{figure}

Figures~\cref{Trans} show the Gibbs free energy, entropy, and heat capacity as functions of the temperature for fixed $r_0=2$. The thermodynamic behavior is parametrized by the horizon radius $r_H$, with
\begin{equation}
G=G(r_H),
\qquad
T=T(r_H),
\qquad
S=S(r_H).
\end{equation}
A thermodynamic critical point associated with the divergence of the heat capacity is determined by
\begin{equation}
\frac{dT}{dr_H}=0,
\end{equation}
or equivalently,
\begin{equation}
C_H\rightarrow\pm\infty.
\end{equation}
Using the thermodynamic identity
\begin{equation}
dG=-S\,dT,
\end{equation}
we have
\begin{equation}
\frac{dG}{dT}=-S,
\qquad
\frac{d^2G}{dT^2}
=-\frac{dS}{dT}
=-\frac{C_H}{T}.
\end{equation}
Therefore, at the critical point, the Gibbs free energy and its first derivative remain finite and continuous, while its second derivative diverges as the heat capacity diverges. The entropy $S=\pi r_H^2$ itself remains smooth as a function of $r_H$, although $dS/dT$ diverges at the temperature extremum because $dT/dr_H=0$.

Accordingly, the divergence of the heat capacity signals a continuous, second-order phase transition within the Ehrenfest classification, provided that the competing thermodynamic branches satisfy the corresponding continuity conditions. In particular, a first-order transition would require a discontinuity in the entropy at the coexistence point and hence a nonzero latent heat. A swallow-tail-like structure in the Gibbs free energy alone is not sufficient to establish a first-order transition; the corresponding competing branches and their thermodynamic dominance must also be examined. In the Schwarzschild limit, $\rho_0=0$, the temperature is monotonic and the heat capacity remains negative, so no such transition occurs.

\section{Summary}

Very recently, a spherically symmetric black hole spacetime surrounded by a King dark matter halo was proposed in Ref.~\cite{Al-Badawi:2026cjx}. However, the metric construction presented therein was not derived as an exact solution of the Einstein field equations. Consequently, the resulting spacetime does not consistently reproduce the prescribed dark matter density through the Einstein tensor, particularly its $G^t_{\ t}$ component. This inconsistency raises the question of whether the physical properties reported in that analysis remain valid when the gravitational field equations are consistently imposed. The main purpose of the present work is therefore to revisit this problem from first principles and construct a self-consistent black hole--dark matter geometry by solving the Einstein equations for the prescribed King halo profile. This approach provides an exact gravitational realization of the King dark matter distribution and allows its physical, optical, dynamical, and thermodynamic properties to be investigated within a consistent spacetime geometry.

We first investigated the motion of photons around a static, spherically symmetric black hole immersed in a King dark matter halo. The dark matter distribution modifies the effective potential for null geodesics and shifts the radius of the circular photon orbit relative to the Schwarzschild value. The corresponding critical impact parameter is also modified, resulting in a change in the apparent black hole shadow. The gravitational deflection of light is likewise affected by the halo, with the dark matter contribution becoming increasingly important in the region where the halo significantly modifies the metric, while the Schwarzschild behavior is recovered asymptotically. These results demonstrate that the King dark matter halo leaves characteristic signatures in the optical properties of the black hole spacetime, which may provide a potential avenue for constraining dark matter environments through black hole shadow and lensing observations.

We next examined the stability of circular photon orbits using the Lyapunov exponent, which characterizes the exponential growth rate of small perturbations around unstable null geodesics. The dark matter halo modifies the magnitude of the Lyapunov exponent and consequently the characteristic instability timescale of the photon orbit. In the eikonal limit, this instability exponent is directly related to the imaginary part of the quasinormal-mode frequencies of massless scalar perturbations. This correspondence establishes a connection between the properties of the unstable photon sphere and the dynamical response of the black hole, linking the classical null-geodesic structure to the damping of high-multipole quasinormal modes.

We also investigated the thermodynamic properties of the black hole--dark matter system by analyzing the horizon structure, Hawking temperature, entropy, heat capacity, and Gibbs free energy. The event horizon is determined by the condition $h(r_H)=0$, and the black hole mass is interpreted as the thermodynamic enthalpy. The Hawking temperature, obtained from the surface gravity at the horizon, depends explicitly on the dark matter parameters $\rho_0$ and $r_0$. The dark matter halo therefore modifies the temperature profile and the resulting thermodynamic behavior. In contrast, the entropy retains the standard Bekenstein--Hawking area law,
\begin{equation}
S=\pi r_H^2=\frac{A}{4},
\end{equation}
and remains positive for $r_H>0$.

The heat capacity $C_H$ characterizes the local thermodynamic stability of the system. Configurations with $C_H>0$ are locally stable, whereas $C_H<0$ corresponds to local thermodynamic instability. Our results show that the presence of the King dark matter halo can generate locally stable branches that are absent in the Schwarzschild limit. The divergences of $C_H$ occur at extrema of the Hawking temperature, where $dT/dr_H=0$, and separate branches with different local stability properties. The extent of these stable and unstable regions depends on the halo parameters $\rho_0$ and $r_0$.

The global thermodynamic behavior was further examined using the Gibbs free energy,
\begin{equation}
G=M-TS.
\end{equation}
Relative to the chosen reference background, configurations with $G<0$ are thermodynamically favored, whereas those with $G>0$ are disfavored. The dark matter halo substantially modifies the Gibbs free energy and can enlarge the parameter region in which the black hole configuration is thermodynamically favored. Thus, the dark matter environment can affect not only the local stability determined by the heat capacity but also the global thermodynamic preference of the black hole configuration.

Finally, we investigated the temperature dependence of the Gibbs free energy, entropy, and heat capacity. The critical behavior is associated with extrema of the temperature and, equivalently, divergences of the heat capacity. Using
\begin{equation}
\frac{dG}{dT}=-S,
\qquad
\frac{d^2G}{dT^2}=-\frac{C_H}{T},
\end{equation}
the divergence of $C_H$ corresponds to a singularity in the second derivative of the Gibbs free energy, while the entropy itself remains continuous. Within the Ehrenfest classification, such behavior is consistent with a continuous, second-order phase transition, provided the corresponding thermodynamic branches satisfy the required continuity conditions. A swallow-tail-like structure in the Gibbs free energy alone is not sufficient to establish a first-order transition; in particular, a first-order transition would require a discontinuity in the entropy and a corresponding nonzero latent heat. In the Schwarzschild limit, $\rho_0=0$, the temperature is monotonic and the heat capacity remains negative, so no analogous phase transition occurs.

\section{Data Availability Statement}
Data sharing not applicable to this article as the current study is purely theoretical.

\bibliography{sn-bibliography}


\end{document}